\documentclass[12pt]{article}
\usepackage[margin=1.5in]{geometry} % Set proper margins
\usepackage{graphicx} % Required for inserting images
\usepackage[breaklinks=true]{hyperref}
\usepackage{xurl} % Allow URL breaks at any character
\usepackage{booktabs} % For professional tables
\usepackage{tabularx} % For flexible column widths
\usepackage{longtable} % For tables that span multiple pages
\usepackage{ltablex} % Combines longtable and tabularx
\usepackage{multirow} % For multi-row cells
\usepackage{array} % For column formatting
\usepackage{tikz} % For taxonomy figure
\usetikzlibrary{shapes.geometric, arrows, positioning, trees}

\title{\textbf{Large Language Models for Power System Applications: A Comprehensive Literature Survey}}
\author{
Muhammad Sarwar$^{1}$, Muhammad Rizwan$^{2}$, Mubushra Aziz$^{2}$, \\
Abdul Rehman Sudais$^{3}$ \\[0.5em]
\footnotesize $^{1}$Dept. of Electrical \& Computer Engineering, Iowa State University, Ames, IA 50010, USA \\
\footnotesize Email: msarwar@iastate.edu \\[0.2em]
\footnotesize $^{2}$Dept. of Computer and Information Sciences, PIEAS, Islamabad, Pakistan \\[0.2em]
\footnotesize $^{3}$Dept. of Computer Sciences, FAST-NUCES, Faisalabad, Pakistan
}
\date{April 2025}

\begin{document}

\maketitle

\begin{abstract}
This comprehensive literature review examines the emerging applications
of Large Language Models (LLMs) in power system engineering. Through a
systematic analysis of recent research published between 2020 and 2025,
we explore how LLMs are being integrated into various aspects of power
system operations, planning, and management. The review covers key
application areas including fault diagnosis, load forecasting,
cybersecurity, control and optimization, system planning, simulation,
and knowledge management. Our findings indicate that while LLMs show
promising potential in enhancing power system operations through their
advanced natural language processing and reasoning capabilities,
significant challenges remain in their practical implementation. These
challenges include limited domain-specific training data, concerns about
reliability and safety in critical infrastructure, and the need for
enhanced explainability. The review also highlights emerging trends such
as the development of power system-specific LLMs and hybrid approaches
combining LLMs with traditional power engineering methods. We identify
crucial research directions for advancing the field, including the
development of specialized architectures, improved security frameworks,
and enhanced integration with existing power system tools. This survey
provides power system researchers and practitioners with a comprehensive
overview of the current state of LLM applications in the field and
outlines future pathways for research and development.
\end{abstract}

\section{Introduction}\label{sec:introduction}

Modern power systems are undergoing a significant transformation driven
by the increasing integration of renewable energy sources, the
proliferation of smart grid technologies, and the growing demands for
enhanced operational efficiency and reliability {[}1{]}, {[}2{]}. This evolution
has led to unprecedented levels of complexity in power grid management,
requiring sophisticated tools and techniques to support intricate
scheduling decisions and ensure system stability {[}1{]}. In this
context, Artificial Intelligence (AI), and particularly Large Language
Models (LLMs), have emerged as promising technologies with the potential
to revolutionize various aspects of power system operation, planning,
and decision-making {[}1{]}. These models, trained on vast amounts of
text data, exhibit remarkable capabilities in natural language
understanding and generation, reasoning, and knowledge retrieval, making
them potentially invaluable assets for navigating the complexities of
the modern power grid {[}4{]}.

This literature review aims to provide a comprehensive survey of the
latest research on the application of LLMs in power system engineering.
The objective is to offer a structured overview of the current research
landscape, critically analyze the methodologies and findings of recent
studies, and identify potential avenues for future research. This review
focuses on studies published within the last three to five years,
identified through a systematic search of Google Scholar, arXiv, and
IEEE Xplore using keywords such as ``Large Language Models power
systems,'' ``LLMs for power grid,'' ``Transformer models in power system
applications,'' and ``Natural Language Processing for power system
analysis.'' The subsequent sections of this report will delve into the
fundamentals of LLMs, provide a detailed survey of their applications
across various power system domains, analyze the methodologies employed
in these studies, identify relevant open-source resources, discuss the
challenges and limitations of using LLMs in this field, and finally,
highlight promising directions for future research. The increasing
reliance on renewable energy and advanced grid technologies has created
a need for more intelligent and adaptive management systems, suggesting
a direct link between the evolution of power systems and the exploration
of LLMs as a solution {[}5{]}. The initial excitement surrounding LLMs
in diverse sectors is now being cautiously extended to the critical
domain of power systems, reflecting a growing trend of investigating
novel technologies while acknowledging the paramount importance of
reliability and safety in grid operations {[}4{]}.

\section{Fundamentals of Large Language Models}\label{sec:fundamentals}

At the core of modern LLMs lies the Transformer architecture, a
groundbreaking neural network design that has significantly advanced the
field of natural language processing {[}4{]}. Unlike previous sequential
models, the Transformer architecture leverages self-attention mechanisms
to understand the relationships between different elements in a
sequence, such as words in a sentence, by allowing for parallel
processing of information {[}4{]}. Key concepts within this architecture
include tokenization, where text is broken down into smaller units
(tokens), and embedding, where these tokens are converted into dense
vector representations that capture their semantic meaning {[}8{]}.
These embeddings, along with positional encodings to maintain the order
of tokens, are then processed through multiple layers of self-attention
and feed-forward networks, enabling the model to learn complex patterns
and dependencies in the input data {[}9{]}.

The capabilities of LLMs extend beyond basic natural language
understanding and generation. Their training on massive datasets allows
them to perform sophisticated reasoning, retrieve vast amounts of
knowledge, and even generate computer code in some cases {[}1{]}. This
versatility makes them potentially valuable tools for a wide range of
power system applications. Different types of LLMs exist, including
encoder-decoder models (e.g., used in translation tasks) and
decoder-only models (e.g., the GPT series), each with its own strengths
and architectural nuances {[}11{]}. Prominent models such as GPT-4
{[}3{]} and Llama {[}6{]} have demonstrated state-of-the-art performance
across various benchmarks, driving their exploration in specialized
domains like power systems. The parallel processing capabilities
inherent in the Transformer architecture offer potential benefits for
handling the large datasets prevalent in power systems, although the
models' limited context window might pose challenges for tasks requiring
the analysis of long temporal dependencies {[}7{]}. Furthermore, the
ability of LLMs to perform few-shot and zero-shot learning is
particularly appealing for power system applications where labeled data
can be scarce or expensive to acquire, potentially revolutionizing how
AI models are developed and deployed in this critical infrastructure
domain {[}7{]}.

\section{Survey of LLM Applications in Power Systems}\label{sec:survey}

\subsection{Fault Diagnosis and Anomaly Detection}\label{subsec:fault-diagnosis}

Recent research has explored the use of LLMs for enhancing fault
diagnosis in power systems. Jing and Rahman {[}13{]} investigated the
application of ChatGPT and GPT-4, combined with advanced prompt
engineering, to improve the accuracy and explainability of fault
diagnoses. Their methodology involved designing comprehensive,
context-aware prompts to guide the LLMs in interpreting complex power
grid data, including real-time sensor readings, historical fault
records, and component descriptions. Experimental results demonstrated
significant improvements in diagnostic accuracy, the quality of
explanations, response coherence, and contextual understanding compared
to baseline prompting methods {[}13{]}. This suggests that carefully
crafted prompts can unlock implicit knowledge within LLMs for specific
power system tasks.

In another study, Matos-Carvalho et al.~{[}17{]} proposed a hybrid deep
learning model for predicting faults in distribution power grid
insulators. Their approach combined an input stage filter for noise
attenuation with an optimized LLM for time series forecasting of leakage
current. The optimized LLM, tuned using tree-structured Parzen
estimation, outperformed state-of-the-art deep learning models in
predicting leakage current increases, indicating the potential of LLMs
in time series forecasting for fault prediction {[}17{]}. The broader
potential of LLMs for anomaly detection in smart grids has also been
highlighted {[}19{]}. These models could be used to analyze various data
sources to identify unusual patterns indicative of faults or other
anomalies, contributing to predictive maintenance strategies {[}5{]}.
Traditional machine learning approaches, such as support vector machines,
have demonstrated success in detecting high impedance faults in distribution
networks {[}6{]}, and LLMs could potentially enhance these methods by
providing more contextual understanding and improved feature extraction
from heterogeneous data sources. Furthermore, in renewable energy systems,
fault ride-through capability remains a critical concern, with techniques
such as superconducting fault current limiters showing promise for
improving system resilience during fault conditions {[}10{]}. LLMs could
assist in optimizing such protection strategies by analyzing operational
patterns and recommending adaptive control actions.

\subsection{Load Forecasting and Demand Response}\label{subsec:load-forecasting}

The application of Natural Language Processing (NLP) techniques,
including those underpinning LLMs, has been explored for improving
electricity demand forecasting. Bai et al.~{[}20{]} investigated the use
of news features, processed through NLP and incorporated into Long
Short-Term Memory (LSTM) networks, to enhance the accuracy of day-ahead
electricity demand forecasts. Their findings indicated that public
sentiment and word vector representations related to transport and
geopolitics had a sustained influence on electricity demand, leading to
improved forecasting performance compared to traditional LSTM models
{[}20{]}. This suggests that incorporating external unstructured data
sources can provide valuable insights for demand forecasting.

The Harvard research team {[}4{]} also highlighted real-time electricity
load forecasting as a potential application area where the strengths of
LLMs could be leveraged. Additionally, L'Heureux et al.~{[}23{]}
proposed a transformer-based architecture for electrical load
forecasting, adapting the NLP transformer workflow to handle time series
data with contextual features. Their results showed that the proposed
approach could successfully handle time series data with contextual
information and outperform state-of-the-art sequence-to-sequence models
{[}23{]}. However, it's important to note that while LLMs are being
explored for load forecasting, some perspectives suggest that existing
AI algorithms might already be well-suited for this task, warranting a
careful evaluation of the added value of LLMs in this specific domain
{[}25{]}.

\subsection{Cybersecurity in Power Systems}\label{subsec:cybersecurity}

The integration of LLMs into power systems presents both opportunities
and significant cybersecurity challenges. Ruan et al.~{[}1{]} conducted
an in-depth analysis of the potential security threats incurred by
applying LLMs to modern power systems. Their work emphasized the need
for urgent research and development of countermeasures to mitigate risks
such as privacy invasion through access to sensitive data, deteriorated
performance due to malicious data manipulation during training or
fine-tuning, semantic divergence leading to incorrect interpretations of
operational states, and denial of service attacks targeting LLM
infrastructure {[}1{]}. The authors highlighted that the very
capabilities that make LLMs useful, such as their broad accessibility
and ability to process diverse data, also make them potential targets
for attackers {[}1{]}.

Beyond identifying threats, research also explores the potential of LLMs
for enhancing cybersecurity in power systems. LLMs could be employed to
analyze network traffic patterns, log files, and threat intelligence
reports to identify potential security threats and suggest mitigation
strategies {[}5{]}. Their natural language processing capabilities could
also be used to analyze textual data for early detection of cyber
threats or attempts to expose sensitive grid infrastructure information
{[}4{]}. This dual-use nature of LLMs in cybersecurity necessitates a
comprehensive understanding of both their potential benefits and the
risks they introduce.

\subsection{Control and Optimization of Power Systems}\label{subsec:control-optimization}

LLMs are being increasingly investigated for their potential to enhance
control and optimization processes in power systems. Zhang {[}26{]}
provided a review summarizing the application of LLM technology in power
system operation and control, particularly in the context of the
increasing penetration of renewable energy sources. The review outlined
the demands placed on AI technology by the evolving power system and
explored the impact of LLMs on various aspects of system management,
including generation, transmission, distribution, consumption, and
equipment {[}26{]}.

Bernier et al.~{[}9{]} introduced SafePowerGraph-LLM, a novel framework
explicitly designed for solving Optimal Power Flow (OPF) problems using
LLMs. The proposed approach combines graph and tabular representations
of power grids to effectively query LLMs, capturing the complex
relationships and constraints within power systems. The framework also
includes tailored in-context learning and fine-tuning protocols for
LLMs, demonstrating reliable performance using off-the-shelf models
{[}9{]}. This work suggests the potential of LLMs to tackle core power
system optimization tasks.

A practical example of LLM application in power system control was
presented in a paper from the International Journal of Innovative
Solutions in Engineering {[}15{]}. The authors demonstrated the
implementation of an LLM agent, based on Llama 3 and executed using
Ollama, for power system control using Python and the Pandapower
library. The LLM agent was able to monitor system conditions, make
decisions, and control aspects of the power system, indicating the
feasibility of using LLMs for real-time control applications {[}15{]}.
Furthermore, Cheng et al.~{[}11{]} introduced GAIA, a pioneering Large
Language Model specifically tailored for power dispatch tasks. They
developed a novel dataset construction technique to fine-tune GAIA for
optimal performance in power dispatch scenarios, demonstrating its
ability to enhance decision-making processes and improve operational
efficiency {[}11{]}. The use of open-source tools in these
implementations highlights a trend towards accessible and reproducible
research in this area. In the context of voltage stability enhancement,
recent research has demonstrated the effectiveness of grid-forming hybrid
PV plants in improving short-term voltage stability {[}14{]}, and the
characterization and mitigation of fault-induced delayed voltage recovery
through dynamic voltage support mechanisms {[}18{]}. These power electronics-based
solutions present opportunities for LLM integration, where language models
could assist in real-time decision support for optimal inverter dispatch
and coordination strategies during voltage disturbances.

\subsection{Power System Planning and Scheduling}\label{subsec:planning-scheduling}

LLMs are also being considered for their role in power system planning
and scheduling. Their ability to process and understand complex
information can be leveraged to assist in generating and analyzing
various scenarios for grid planning and expansion, taking into account
factors such as load growth, renewable integration, and regulatory
changes {[}5{]}. Mongaillard et al.~{[}27{]} proposed a user-centric
approach for power scheduling using LLM agents. Their novel architecture
constructs three LLM agents to convert arbitrary user voice requests
into resource allocation vectors, aiming to create more flexible and
personalized energy services, particularly in wireless and digitalized
energy networks {[}27{]}. The capability of LLMs to optimize scheduling
decisions within the ever-expanding operational scope of modern power
systems has also been noted {[}1{]}. This suggests a potential shift
towards more user-responsive and intelligent grid operations.

\subsection{Simulation and Modeling of Power Systems}\label{subsec:simulation}

Given the heavy reliance on simulations in power system research and
operation, enabling LLMs to perform these simulations is a significant
area of interest. Jia et al.~{[}16{]} proposed a modular framework to
enhance the ability of LLMs to perform power system simulations using
previously unseen tools. Their framework integrates expertise from both
the power system and LLM domains, incorporating techniques such as
prompt engineering, retrieval-augmented generation, toolbox refinement,
and a feedback loop for error correction {[}16{]}. Validated using the
DALINE toolbox, their approach significantly improved the simulation
coding accuracy of LLMs, highlighting their potential as research
assistants in power systems {[}16{]}.

However, LLMs face significant challenges in power system simulations
due to their limited pre-existing knowledge and the inherent complexity
of power grids {[}1{]}. Despite these challenges, the concept of LLMs
acting as research assistants capable of performing simulations holds
considerable promise for advancing power system studies by automating
laborious tasks and allowing human researchers to focus on higher-level
design and interpretation {[}16{]}. The development of specialized
frameworks is crucial for overcoming the limitations of off-the-shelf
LLMs in this domain.

\subsection{Knowledge Management and Decision Support for Power Engineers}\label{subsec:knowledge-management}

The vast amount of technical documentation, operational procedures, and
real-time data in power systems presents a significant challenge for
knowledge management. LLMs offer the potential to enhance the extraction
of critical information from these datasets, acting as intelligent data
retrieval and question-answering systems for power system operators
{[}1{]}. Furthermore, LLMs can improve human-computer interactions
within power systems by providing intuitive presentations of complex
data and operational states, supporting operators in making
well-informed decisions {[}1{]}.

Recent research has also explored the use of LLMs in developing
domain-specific lexicons for urban power grid design. Xu et al.~{[}29{]}
proposed a framework leveraging LLMs for multi-level term extraction and
synonym expansion to address the challenges of terminology updates and
semantic ambiguity in this field. Their approach successfully
constructed a high-precision terminology dictionary, improving the
semantic parsing capabilities of intelligent design systems {[}29{]}.
This demonstrates the potential of LLMs to enhance knowledge management
and decision support tools for power engineers.

\subsection{Other Emerging Applications}\label{subsec:emerging-applications}

Beyond the core application areas, LLMs are being explored for other
innovative uses in power systems. ChatGrid, for instance, is a
generative AI tool developed for power grid visualization, utilizing a
publicly available large language model to interpret user queries and
generate Structured Query Language (SQL) for searching internal grid
data {[}30{]}. The emergence of foundation models specifically for the
power and energy sectors, such as PowerPM {[}31{]} and RE-LLaMA
{[}32{]}, indicates a trend towards creating AI systems with a deeper
understanding of the unique characteristics and data of these domains,
potentially leading to more specialized and effective applications.
PowerPM aims to model electricity time series data for various
downstream tasks {[}31{]}, while RE-LLaMA is focused on renewable and
hydrogen energy deployment strategies {[}32{]}. These developments
suggest a growing recognition of the need for domain-specific LLMs in
the power industry.

% Summary Table of LLM Applications (using longtable for page breaks)
\begin{small}
\begin{longtable}{>{\raggedright\arraybackslash}p{2.2cm}>{\raggedright\arraybackslash}p{4.5cm}>{\raggedright\arraybackslash}p{3.5cm}>{\raggedright\arraybackslash}p{1.8cm}}
\caption{Summary of LLM Applications in Power Systems} \label{tab:llm-applications} \\
\toprule
\textbf{Application Domain} & \textbf{Key Use Cases} & \textbf{LLM Techniques} & \textbf{Maturity} \\
\midrule
\endfirsthead
\multicolumn{4}{c}{\tablename\ \thetable{} -- continued from previous page} \\
\toprule
\textbf{Application Domain} & \textbf{Key Use Cases} & \textbf{LLM Techniques} & \textbf{Maturity} \\
\midrule
\endhead
\midrule
\multicolumn{4}{r}{Continued on next page...} \\
\endfoot
\bottomrule
\endlastfoot
Fault Diagnosis \& Anomaly Detection & Fault classification, leakage current prediction, anomaly identification, predictive maintenance & Prompt engineering, time series forecasting, hybrid ML-LLM models & Medium \\
\addlinespace
Load Forecasting \& Demand Response & Day-ahead demand prediction, real-time load forecasting, sentiment-based forecasting & Transformer architectures, NLP feature extraction, LSTM integration & Medium-High \\
\addlinespace
Cybersecurity & Threat detection, log analysis, vulnerability assessment, attack mitigation & Text analysis, pattern recognition, threat intelligence processing & Low-Medium \\
\addlinespace
Control \& Optimization & Optimal power flow, power dispatch, real-time control, voltage stability support & Graph-based representations, LoRA fine-tuning, agent-based systems & Medium \\
\addlinespace
Planning \& Scheduling & Scenario analysis, resource allocation, user-centric scheduling & Multi-agent LLM systems, voice-to-action conversion & Low \\
\addlinespace
Simulation \& Modeling & Automated simulation coding, tool integration, research assistance & RAG, prompt engineering, feedback loops & Low-Medium \\
\addlinespace
Knowledge Management & Information extraction, Q\&A systems, terminology dictionaries, decision support & Semantic parsing, lexicon building, document analysis & Medium \\
\addlinespace
Emerging Applications & Grid visualization, foundation models (PowerPM, RE-LLaMA), SQL generation & Domain-specific pre-training, multi-modal processing & Low \\
\end{longtable}
\end{small}

% Taxonomy Figure
\begin{figure}[htbp]
\centering
\resizebox{\textwidth}{!}{%
\begin{tikzpicture}[
    level 1/.style={sibling distance=5cm, level distance=2.2cm},
    level 2/.style={sibling distance=2.8cm, level distance=2cm},
    every node/.style={rectangle, draw, rounded corners, align=center, font=\small, minimum height=0.8cm, minimum width=1.8cm},
    root/.style={fill=blue!20, font=\small\bfseries, minimum width=3.5cm, minimum height=1cm},
    cat1/.style={fill=green!20, minimum width=2.2cm},
    cat2/.style={fill=orange!20, minimum width=2.2cm},
    cat3/.style={fill=purple!20, minimum width=2.2cm},
    leaf/.style={fill=gray!10, font=\footnotesize, minimum width=2cm, minimum height=0.9cm}
]
\node[root] {LLMs in\\Power Systems}
    child {node[cat1] {Operations}
        child {node[leaf] {Fault\\Diagnosis}}
        child {node[leaf] {Load\\Forecasting}}
        child {node[leaf] {Control \&\\Optimization}}
    }
    child {node[cat2] {Planning}
        child {node[leaf] {Scheduling}}
        child {node[leaf] {Simulation}}
        child {node[leaf] {Grid\\Expansion}}
    }
    child {node[cat3] {Support}
        child {node[leaf] {Knowledge\\Management}}
        child {node[leaf] {Cyber-\\security}}
        child {node[leaf] {Decision\\Support}}
    };
\end{tikzpicture}%
}
\caption{Taxonomy of LLM Applications in Power Systems}
\label{fig:taxonomy}
\end{figure}
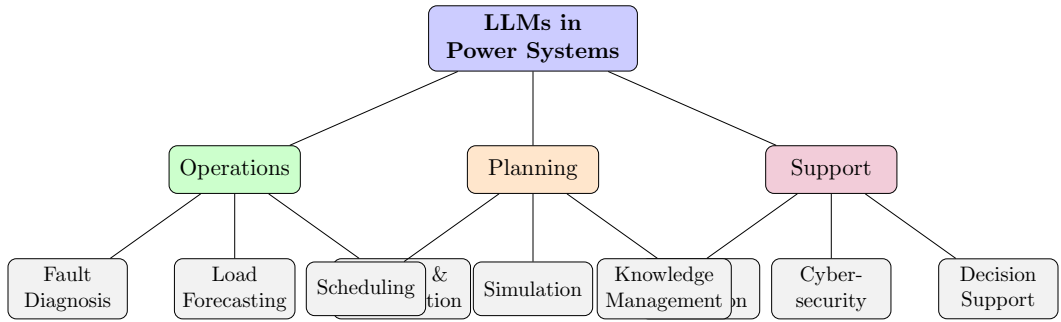

\section{Methodological Analysis Across Applications}\label{sec:methodology}

The surveyed applications of LLMs in power systems utilize a variety of
LLM architectures and fine-tuning approaches. Models like GPT-4 {[}3{]}
and Llama {[}3{]} are frequently employed, often with specific
fine-tuning techniques. For instance, Bernier et al.~{[}9{]} utilized
Low-Rank Adaptation (LoRA) for efficient fine-tuning of LLMs for OPF
problems. The choice of LLM often depends on the specific task and the
availability of computational resources {[}9{]}.

Datasets used in these studies vary widely in size, source, and
characteristics. Some research utilizes simulated data generated from
power system models {[}33{]}, while others employ real-world data such
as sensor readings {[}13{]}, electricity demand data {[}21{]}, or news
articles {[}21{]}. The challenge of imbalanced datasets, where the
occurrence of certain events (e.g., faults) is rare, is also addressed
in some studies through techniques like synthetic data generation using
conditional Wasserstein generative adversarial networks {[}33{]}. The
availability of high-quality, domain-specific datasets remains a
critical factor for the successful training and evaluation of LLMs in
this field {[}7{]}.

Performance metrics also vary depending on the application. For fault
diagnosis and anomaly detection, accuracy and explainability quality are
often key metrics {[}13{]}. In load forecasting, Root Mean Squared Error
(RMSE), Mean Absolute Error (MAE), and other statistical measures are
commonly used {[}21{]}. For optimization tasks like OPF, the error in
estimating power flow variables is a crucial metric {[}9{]}. In
simulation tasks, coding accuracy is used to evaluate the LLM's ability
to generate correct simulation scripts {[}16{]}. The diverse range of
metrics reflects the multifaceted nature of LLM applications in power
systems.

\section{Open-Source Resources and Datasets}\label{sec:resources}

Several open-source resources and datasets are mentioned across the
surveyed literature. Power system simulation tools and libraries such as
Pandapower {[}15{]}, OpenDSS {[}16{]}, DALINE {[}16{]}, and MATPOWER
{[}34{]} are utilized in various studies for modeling and simulating
power systems. Open-source LLM models like Llama {[}3{]} are also
employed, facilitating accessible research in this area. While specific
publicly available power system datasets are not consistently mentioned,
some studies utilize publicly accessible data sources like news articles
from the BBC {[}21{]} or electricity demand data from ENTSO-E {[}21{]}.
Mentions of open-source code repositories, such as GitHub links for
weather and climate foundation models {[}35{]} and LLM reasoning
benchmarks {[}36{]}, indicate a growing trend towards sharing code and
resources within the research community. The reliance on open-source
tools and models promotes transparency and reproducibility, which are
essential for the advancement of LLM applications in power systems. The
availability of diverse datasets, both simulated and real-world, is
crucial for training and evaluating LLMs for a wide range of power
system applications, highlighting the importance of continued efforts to
create and share such resources.

\section{Challenges, Limitations, and Potential Security Threats}\label{sec:challenges}

Despite the promising applications, several challenges and limitations
hinder the widespread adoption of LLMs in power systems. A significant
challenge is the limited availability of domain-specific data for
pre-training and fine-tuning LLMs due to privacy concerns and regulatory
restrictions {[}7{]}. This scarcity of data can limit the models'
ability to accurately understand and reason about complex power system
operations. General-purpose LLMs also suffer from issues like generative
stochasticity, inconsistency, and the potential for hallucinations
(generating factually incorrect information), which can be particularly
problematic for safety-critical applications in power systems {[}4{]}.
Their limitations in complex mathematical reasoning and handling of
highly specialized domain knowledge further restrict their applicability
in certain power engineering tasks {[}4{]}.

Moreover, the integration of LLMs into power systems introduces
potential security threats. These include the risk of privacy invasion
through access to sensitive operational data, the possibility of
deteriorated performance if LLMs are trained or fine-tuned on
maliciously altered data, semantic divergence leading to incorrect
interpretations of system states, and the potential for denial of
service attacks targeting the LLM infrastructure itself {[}1{]}. The
high computational and energy costs associated with training and
deploying large LLMs also pose a challenge, raising concerns about the
sustainability of their widespread use {[}38{]}. Finally, the lack of
explainability and transparency in the decision-making processes of many
LLMs makes it difficult to build trust among power system operators and
engineers, which is crucial for their adoption in critical
infrastructure {[}13{]}. The inherent limitations related to
reliability, safety, and domain-specific reasoning present significant
barriers to the direct deployment of LLMs in core power system
operational roles, necessitating a careful and phased approach to their
integration {[}4{]}.

\section{Future Research Directions}\label{sec:future}

Future research in the application of LLMs in power systems should focus
on several key areas to address the current limitations and unlock their
full potential. Developing hybrid AI systems that combine the strengths
of LLMs with traditional power system analysis and optimization
techniques is a promising direction {[}7{]}. Investigating the use of
reinforcement learning to enable LLM agents to learn and improve their
decision-making capabilities in power system control could also lead to
significant advancements {[}3{]}. Recent work on deep reinforcement learning
frameworks for short-term voltage stability improvement {[}24{]} demonstrates
the potential of combining learning-based approaches with power system control,
suggesting that hybrid LLM-reinforcement learning architectures could yield
powerful tools for adaptive grid management. Further research on Retrieval
Augmented Generation (RAG) techniques is crucial for enhancing the
accuracy and reliability of LLM responses by grounding them in
domain-specific knowledge and real-time data {[}7{]}.

Exploring the potential of multi-agent systems where multiple LLM agents
collaborate to manage complex power system tasks could offer new
solutions for distributed control and optimization {[}7{]}. Developing
robust security frameworks and countermeasures to address the potential
threats introduced by LLMs in power systems is paramount for their safe
and reliable deployment {[}1{]}. Efforts should also be directed towards
improving the explainability and transparency of LLM-based solutions to
build trust among power system operators and engineers {[}7{]}.
Investigating the application of multi-modal LLMs that can process and
reason with different types of power system data, such as time series,
images, and diagrams, could lead to more comprehensive and insightful
applications {[}1{]}. Finally, exploring the development of specialized,
smaller-scale LLMs pre-trained on extensive power system-specific data
could enhance domain expertise while reducing computational costs and
energy consumption {[}29{]}. Overcoming current limitations by focusing
on these research directions will be essential for creating valuable and
trustworthy LLM-based tools for the power systems domain.

\section{Conclusion}\label{sec:conclusion}

This literature review has provided a comprehensive survey of recent
research on the application of Large Language Models in power system
engineering. The findings highlight a growing interest in leveraging the
capabilities of LLMs across various domains, including fault diagnosis,
load forecasting, cybersecurity, control and optimization, planning and
scheduling, simulation and modeling, and knowledge management. The
potential of LLMs to enhance efficiency, reliability, and
decision-making in power systems is evident in the diverse range of
applications being explored. However, the review also underscores
significant challenges and limitations that need to be addressed for the
successful integration of LLMs into this critical infrastructure sector.
These include the scarcity of domain-specific data, issues with the
reliability and safety of LLM outputs, limitations in reasoning and
mathematical abilities, potential security threats, and the need for
explainability and trust. Future research should focus on developing
hybrid approaches, enhancing domain-specific knowledge, ensuring
security and trustworthiness, and improving the transparency of
LLM-based solutions. Addressing these challenges and pursuing the
promising future research directions outlined in this report will be
crucial for realizing the transformative potential of LLMs in shaping
the next generation of intelligent and resilient power systems.

\section*{References}\label{sec:references}

\begin{enumerate}
\def\labelenumi{\arabic{enumi}.}

\item
  J. Ruan, H. Wang, Y. Wang, and H. Chen, ``Applying Large Language Models to Power Systems: Potential Security
  Threats,'' arXiv preprint arXiv:2311.13361, 2023.
  \url{https://arxiv.org/abs/2311.13361}
\item
  M. Sarwar and B. Asad, ``A Review on Future Power Systems; Technologies
  and Research for Smart Grids,'' International Conference on Emerging
  Technologies (ICET), Islamabad, Pakistan, 2017.
\item
  X. Chen et al., ``Large Foundation Models for Power Systems,'' arXiv preprint arXiv:2312.07044, 2023.
  \url{https://arxiv.org/abs/2312.07044}
\item
  Harvard Utility Analytics, ``Large Language Models: Applications, Limitations and Potential Risks
  for Power Grids,'' 2025.
  \url{https://utilityanalytics.com/large-language-models-grid-analytics/}
\item
  ``Large Language Models in Power Systems: Enhancing Control and
  Decision-Making,'' International Journal of Innovative Solutions in Engineering, 2025.
  \url{https://ijise.ba/wp-content/uploads/2025/01/Vol.-1-No.-1-Article-2.pdf}
\item
  M. Sarwar, F. Mehmood, M. Abid, A. Q. Khan, S. T. Gul, and A. S. Khan,
  ``High impedance fault detection and isolation in power distribution
  networks using support vector machines,'' Journal of King Saud
  University-Engineering Sciences, vol. 32, no. 8, pp. 524-535, 2020.
\item
  A. Eberle et al., ``Exploring the Capabilities and Limitations of Large Language Models in
  the Electric Energy Sector,'' arXiv preprint arXiv:2403.09125, 2024.
  \url{https://arxiv.org/abs/2403.09125}
\item
  J. Lago et al., ``A Transformer approach for Electricity Price Forecasting,'' arXiv preprint arXiv:2403.16108, 2024.
  \url{https://arxiv.org/abs/2403.16108}
\item
  N. Bernier et al., ``SafePowerGraph-LLM: Novel Power Grid Graph Embedding and Optimization
  with Large Language Models,'' arXiv preprint arXiv:2501.07639, 2025.
  \url{https://arxiv.org/abs/2501.07639}
\item
  M. Sarwar, B. Hussain, A. Hussain, and M. Abubakar, ``Improvement of
  Fault Ride Through Capability of DFIG-based Wind Turbine Systems Using
  Superconducting Fault Current Limiter,'' 2019 IEEE Innovative Smart
  Grid Technologies - Asia (ISGT Asia), Chengdu, China, 2019.
\item
  Y. Cheng et al., ``GAIA -- A Large Language Model for Advanced Power Dispatch,'' arXiv preprint arXiv:2408.03847, 2024.
  \url{https://arxiv.org/abs/2408.03847}
\item
  S. Thirunavukarasu et al., ``Leveraging Generative AI and Large Language Models: A Comprehensive
  Roadmap for Healthcare Integration,'' PMC, 2023.
  \url{https://pmc.ncbi.nlm.nih.gov/articles/PMC10606429/}
\item
  Y. Jing and S. Rahman, ``Enhancing Power System Fault Diagnosis with ChatGPT,'' arXiv preprint arXiv:2407.08836, 2024.
  \url{https://arxiv.org/abs/2407.08836}
\item
  M. Sarwar, V. Ajjarapu, A. R. R. Matavalam, S. Roy, and H. Villegas-Pico,
  ``Short-term Voltage Stability Improvement in Power System through
  Grid-forming Hybrid PV Plants,'' IEEE Transactions on Industry
  Applications, 2025.
\item
  ``Large Language Models in Power Systems: Enhancing Control and
  Decision-Making,'' ResearchGate, 2025.
  \url{https://www.researchgate.net/publication/388126751}
\item
  M. Jia et al., ``Enabling Large Language Models to Perform Power System Simulations
  with Previously Unseen Tools: A Case of Daline,'' arXiv preprint arXiv:2406.17215, 2024.
  \url{https://arxiv.org/abs/2406.17215}
\item
  J. P. Matos-Carvalho et al., ``Time series forecasting based on optimized LLM for
  fault prediction in distribution power grid insulators,'' arXiv preprint arXiv:2502.17341, 2025.
  \url{https://arxiv.org/abs/2502.17341}
\item
  M. Sarwar, A. R. R. Matavalam, and V. Ajjarapu, ``Characterization and
  Mitigation of Fault Induced Delayed Voltage Recovery with Dynamic
  Voltage Support by Hybrid PV Plants,'' 2022 North American Power
  Symposium, Utah, USA, pp. 1-6, 2022.
\item
  Z. Liu et al., ``Risks of Practicing Large Language Models in Smart Grid: Threat
  Modeling and Validation,'' arXiv preprint arXiv:2405.06237, 2024.
  \url{https://arxiv.org/abs/2405.06237}
\item
  Y. Bai et al., ``News and Load: A Quantitative Exploration of Natural
  Language Processing Applications for Forecasting Day-ahead Electricity
  System Demand,'' arXiv preprint arXiv:2301.07535, 2023.
  \url{https://arxiv.org/abs/2301.07535}
\item
  Y. Bai et al., ``Electricity Demand Forecasting through Natural Language Processing
  with Long Short-Term Memory Networks,'' arXiv preprint arXiv:2309.06793, 2023.
  \url{https://arxiv.org/abs/2309.06793}
\item
  ``Research Examines Potential Role of Large Language Models in Grid
  Management,'' Public Power Magazine, 2025.
  \url{https://www.publicpower.org/periodical/article/research-examines-potential-role-large-language-models-grid-management}
\item
  A. L'Heureux et al., ``Transformer-Based Model for Electrical Load Forecasting,'' Energies, vol. 15, no. 14, 2022.
  \url{https://www.mdpi.com/1996-1073/15/14/4993}
\item
  M. Sarwar, A. R. R. Matavalam, and V. Ajjarapu, ``Deep Reinforcement
  Learning Framework for Short-Term Voltage Stability Improvement,''
  2023 Texas Power and Energy Conference, 2023.
\item
  ``What are possible roles of LLM in the research of low-carbon energy
  systems?'' ResearchGate Discussion, 2024.
  \url{https://www.researchgate.net/post/What_are_possibles_roles_of_LLM_in_the_research_of_low-carbon_energy_systems}
\item
  L. Zhang, ``Application of Large Language Models in Power System Operation,'' Journal of Computing and Electronic Information Management, 2024.
  \url{https://drpress.org/ojs/index.php/jceim/article/view/28339}
\item
  G. Mongaillard et al., ``Large Language Models for Power Scheduling: A
  User-Centric Approach,'' arXiv preprint arXiv:2407.00476, 2024.
  \url{https://arxiv.org/abs/2407.00476}
\item
  M. Jia et al., ``Enhancing LLMs for Power System Simulations: A Feedback-driven
  Multi-agent Framework,'' arXiv preprint arXiv:2411.16707, 2024.
  \url{https://arxiv.org/abs/2411.16707}
\item
  Y. Xu et al., ``LLM-Enhanced Framework for Building Domain-Specific Lexicon for
  Urban Power Grid Design,'' Applied Sciences, vol. 15, no. 8, 2025.
  \url{https://www.mdpi.com/2076-3417/15/8/4134}
\item
  ``ChatGrid: A New Generative AI Tool for Power Grid Visualization,'' Pacific Northwest National Laboratory News Release, 2024.
  \url{https://www.pnnl.gov/news-media/chatgridtm-new-generative-ai-tool-power-grid-visualization}
\item
  S. Liang et al., ``PowerPM: Foundation Model for Power Systems,'' arXiv preprint arXiv:2408.04057, 2024.
  \url{https://arxiv.org/abs/2408.04057}
\item
  K. Zheng et al., ``Domain-Specific Large Language Model for Renewable Energy and
  Hydrogen Energy Deployment,'' Energies, vol. 17, no. 23, 2024.
  \url{https://www.mdpi.com/1996-1073/17/23/6063}
\item
  Z. Li et al., ``Deep learning based on Transformer architecture for power system
  short-term voltage stability assessment with class imbalance,'' arXiv preprint arXiv:2310.11690, 2023.
  \url{https://arxiv.org/abs/2310.11690}
\item
  Y. Wang et al., ``Powerformer: A Section-adaptive Transformer for Power Flow Adjustment,'' arXiv preprint arXiv:2401.02771, 2024.
  \url{https://arxiv.org/abs/2401.02771}
\item
  ``Awesome-Foundation-Models-for-Weather-and-Climate,'' GitHub Repository, 2024.
  \url{https://github.com/shengchaochen82/Awesome-Foundation-Models-for-Weather-and-Climate}
\item
  ``LLM-arxiv-daily: Automatically update arXiv papers about LLM
  Reasoning, LLM Evaluation, and MLLM,'' GitHub Repository, 2024.
  \url{https://github.com/Xuchen-Li/llm-arxiv-daily}
\item
  M. K. Heris, ``Powering the Grid with Language: How LLMs Are Transforming Energy
  Systems,'' Medium, 2025.
  \url{https://kalami.medium.com/powering-the-grid-with-language-how-llms-are-transforming-energy-systems-54a12a915323}
\item
  P. Sahu et al., ``The Unseen AI Disruptions for Power Grids: LLM-Induced Transients,'' arXiv preprint arXiv:2409.11416, 2024.
  \url{https://arxiv.org/abs/2409.11416}
\item
  M. Wang et al., ``An Efficient and Explainable Transformer-Based
  Few-Shot Learning for Modeling Electricity Consumption Profiles Across
  Thousands of Domains,'' arXiv preprint arXiv:2408.08399, 2024.
  \url{https://arxiv.org/abs/2408.08399}
\end{enumerate}

\end{document}